\documentclass[aps,letterpaper, floatfix,twocolumn,superscriptaddress]{revtex4-1}
\usepackage{setspace,comment,ulem}

\usepackage{color}
\usepackage{braket}
\usepackage{amsmath, stmaryrd,amssymb}
\usepackage[pdftex]{graphicx}

\newcommand{\be}{\begin{equation}}
\newcommand{\ee}{\end{equation}}
\def\ba{\begin{array}}
\def\ea{\end{array}}
\def\sx{\sigma^x}
\def\sz{\sigma^z}
\def\rr{\right}
\def\l{\left}
\renewcommand{\H}{{\cal H}}
\newcommand{\summ}{\sum}

\begin{document}
\title{The Hilbert-glass transition: new universality of temperature-tuned many-body dynamical quantum criticality}
\author{David Pekker}
\affiliation{Department of Physics, California Institute of
  Technology, Pasadena, CA 91125, USA}
  \author{Gil Refael}
\affiliation{Department of Physics, California Institute of
  Technology, Pasadena, CA 91125, USA}
\author{Ehud Altman}
\affiliation{Department of Condensed Matter Physics, Weizmann Institute of Science, Rehovot 76100, Israel}
\affiliation{Department of Physics, University of California, Berkeley, CA 94720, USA}
\author{Eugene Demler}
\affiliation{
Department of Physics, Harvard University, Cambridge MA 02138, USA}
\author{Vadim Oganesyan}
\affiliation{Department of Engineering Science and Physics,
College of Staten Island, CUNY, Staten Island, NY 10314, USA}
\affiliation{The Graduate Center, CUNY, New York, NY 10016, USA}

\begin{abstract}
We consider a new class of unconventional critical phenomena that is characterized by singularities only in dynamical quantities and has no thermodynamic signatures. A possible example is the recently proposed many-body localization transition, in which transport coefficients vanish at a critical temperature. Describing this unconventional quantum criticality has been technically challenging as understanding the finite-temperature dynamics requires the knowledge of a large number of many-body eigenstates. Here we develop a real-space renormalization group method for excited state (RSRG-X),  that allow us to overcome this challenge, and establish the existence and universal properties of such temperature-tuned dynamical phase transitions. We characterize a specific example: the 1D disordered transverse field Ising model with interactions. Using RSRG-X, we find a finite-temperature transition, between two localized phases, characterized by non-analyticities of the dynamic spin correlation function and the low frequency heat conductivity.

\end{abstract}
\maketitle

\section{Introduction}
Our ability to describe emergent behavior in many-body systems relies, to a large extent, on 
the universality of critical phenomena associated with phase transitions and spontaneous symmetry breaking. Spontaneous symmetry breaking plays an important role even in disordered systems. For example, the spin-glass transition in classical magnets with random interactions follows this paradigm: as temperature drops, a specific frozen magnetization pattern which breaks an Ising symmetry emerges~\cite{Fisher1986,Fisher1988}. In one dimension, however, there are strong arguments, which forbid spontaneous symmetry breaking and more generally thermodynamic phase transitions from occurring at any non-vanishing temperature~\cite{VanHove1950,Landau1959}.

Do these arguments rule out the observation of critical phenomena in one dimensional systems at non-vanishing temperature?
A recent theoretical work, which generalizes the phenomenon of Anderson localization~\cite{Anderson1958} to interacting many-body systems, suggests otherwise~\cite{Basko2006,Basko2007}. Its intriguing prediction is that an isolated many-body system subject to strong disorder can undergo a phase transition, from a state with strictly zero  (thermal) conductivity at low temperature to a metallic phase above a critical temperature. This transition has only dynamical manifestations and no thermodynamic ones, and is, in this sense, a many-body extension of the mobility edge~\cite{Mott1987} in the Anderson localization transition in 3D. A similar many-body transition was also suggested to exist in the 1D bosonic
case~\cite{Aleiner2010}, at infinite temperatures as a function of
disorder strength~\cite{Oganesyan2007,Monthus2010,Pal2010}, and in quasiperiodic systems without disorder~\cite{Iyer2013}. Very little,
however, is known about the universality of such transitions.

\begin{figure*}
\includegraphics[width=\textwidth]{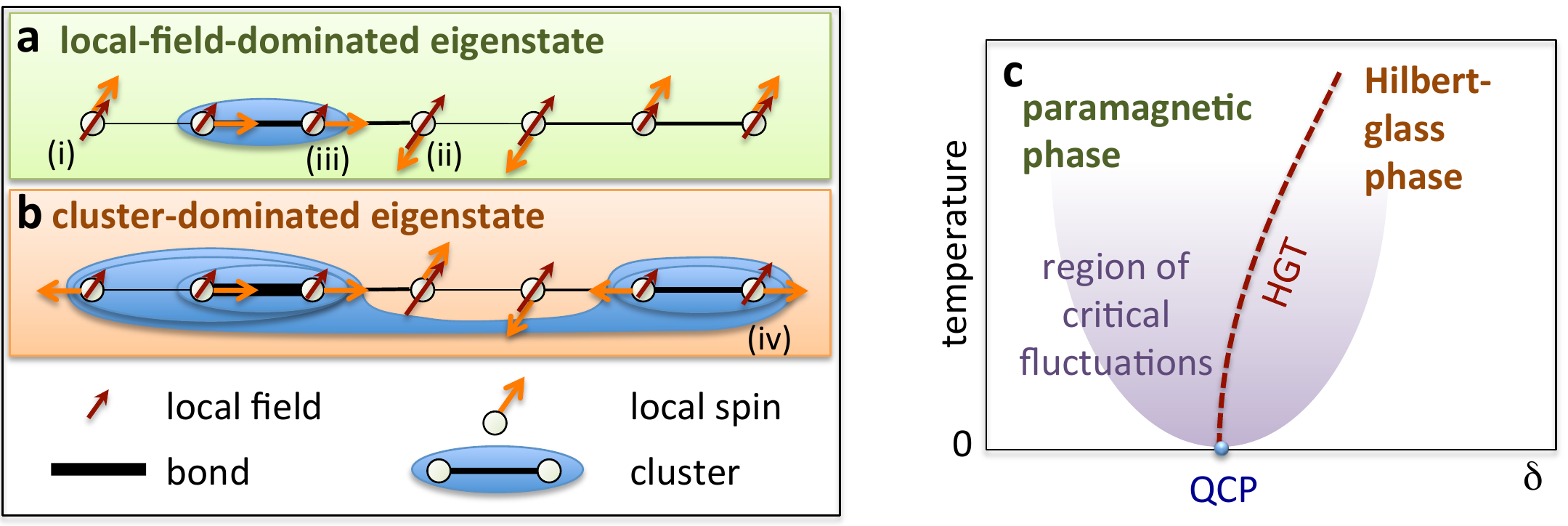}
\caption{
{\bf a} Schematic representation of a typical eigenstate in the paramagnetic (local-field-dominated) phase: the local spins are largely aligned or anti-aligned with the local fields; rare clusters can still be present.
{\bf b} Hilbert-glass (cluster-dominated) eigenstate: local spins form magnetic clusters, that often contain domain walls; rare isolated spins can still be present. 
{\bf c} Schematic phase diagram of the Hilbert-glass transition (HGT) in the tuning parameter--temperature plane. The diagram shows the HGT line separating the paramagnetic (local-field-dominated) and the Hilbert-glass (cluster-dominated) phases terminating in a zero-temperature quantum critical point (QCP). The behavior of equilibrium thermodynamic observables at finite temperatures is governed by the QCP in the region of critical fluctuations~\cite{Sachdev2011}. While equilibrium thermodynamic observables have no singularities in the phase diagram, with the exception of the zero temperature QCP, dynamical observables show singular behavior along the HGT line. 
\label{fig:one}
}
\end{figure*}

In this paper we uncover a wider class of unconventional critical
  phenomena, whose existence was hinted by the many-body
  localization transition example. By investigating a random spin
chain with generic interactions we find and characterize another
instance of this new class of transitions that exhibit non-analytic
dynamical behavior as a function of temperature, but lack any strictly
thermodynamic signatures. The transition stems from the structure of
the many-body eigenstates, which decidedly changes as we tune the
extensive many-body energy~\cite{Huse2013}. On one side of the
transition (the paramagnetic side) the eigenstates consist of most
spins aligned or anti-aligned with the random local field, while on
the other side (the glass side) spins show a pattern of frozen
magnetization that breaks a $\mathbb{Z}_2$ symmetry (See
Fig. \ref{fig:one}a and b). The magnetization pattern is random,
however, and varies from eigenstate to eigenstate as if each is a
classical spin-glass ground state of a different disorder
realization. The signatures of the phases and phase transition appear not
in a single state, but rather in a large
portion of the many body Hilbert space. Therefore, we coin the term Hilbert-glass transition (HGT) to described this type of dynamical quantum criticality. The properties of the HGT are summarized in the generic phase diagram Fig.~\ref{fig:one}c.

The challenges for analyzing the HGT are clear: a complete knowledge of the properties of an entire excitation spectrum -- wave functions and energies alike -- is necessary. Previous efforts to characterize quantum dynamical transitions, such as the many-body localization transition, relied on Exact Diagonalization of small systems~\cite{Oganesyan2007,Karahalios2009,Pal2010,Barisic2010,Berkelbach2010,Khatami2012,Kim2013,Serbyn2013,Serbyn2013a}. Here, we pursue an alternative approach based on the strong disorder real-space renormalization group (RSRG) method~\cite{Dasgupta1980, Bhatt1982, Fisher1992, Fisher1994, Fisher1995, Motrunich2001} and its extensions to unitary time evolution~\cite{Vosk2013}.  We generalize the method to deal with arbitrary-energy excitations, hence the acronym RSRG-X. The RSRG-X allows us to investigate the dynamics of an arbitrary-temperature thermal state of strongly disordered systems containing thousands of sites, i.e., accessing systems two orders of magnitude larger than what exact diagonalization can access.

Unlike the many-body localization transition, the transition we describe occurs between two localized phases, which therefore cannot thermalize on their own~\cite{Popescu2006,Rigol2008,Zangara2013}. Nonetheless there are physical settings in which it is meaningful to discuss thermal response functions and temperature tuned transitions in these systems. First, one can imagine preparing the system in equilibrium by connecting it to an external bath, which is adiabatically disconnected before the start of the response measurement. Alternatively, the system could be weakly coupled to a thermal bath at all times. In the second case there is a time scale, set by the bath coupling strength, beyond which the bath dominates the dynamics. The strong-disorder fixed point that we describe would provide a faithful description of the dynamical behavior on shorter time scales. 

The model we investigate with the RSRG-X is the generalized quantum Ising model 
\begin{align}
\H_\text{hJJ'}=-\sum_i \left( J_i \sx_i \sx_{i+1} + h_i \sz_i + J'_i \sz_i \sz_{i+1} \right). \label{hjjp}
\end{align}
Without the last $J'$ term Eq.~\eqref{hjjp} represents the usual transverse field Ising model (TFIM), which was the  first arena in which real-space renormalization group was used to elucidate the novel universal properties of phase transition dominated by strong disorder~\cite{Fisher1992,Fisher1995}.  In particular, the RSRG analysis yields the infinite randomness energy-length scaling $\log (1/E)\sim \ell^{1/2}$, in contrast to the $E\sim \ell^{-z}$ scaling in conventional critical systems. 

The TFIM makes a natural starting point for investigating dynamical
critical points in strongly disordered systems. The TFIM, Eq.
(\ref{hjjp}), however, can be mapped to a free fermion theory, making its
dynamics fundamentally equivalent to that of a certain class of single
particle Anderson localization (with particle-hole symmetry). We avoid
this problem by adding the $J'$ interaction term, which preserves the
$\mathbb{Z}_2$ symmetry, while making the model intrinsically interacting.
It has been shown that the dynamics in the many-body localized phase, i.e.
in presence of interactions, can be different than in the non-interacting
case~\cite{Bardarson2012,Vosk2013,Serbyn2013a,Serbyn2013,Kim2013}.

The paper is organized as follows. In Sec.~\ref{sec:transition} we develop the RSRG-X procedure and consider the flows it produces. The flows reveal the evolution of the many-body eigenstate structure as temperature is varied, which allows us to identify the dynamical phase transition and construct the phase diagram. We find that the temperature-tuned transition is controlled by an infinite-randomness critical point, with the same scaling properties as the zero temperature quantum phase transition~\cite{Fisher1992}. In Sec.~\ref{sec:autocorrelation} and \ref{sec:sigmaT} we examine two dynamical observables: the low frequency spin autocorrelation function (and an associated Edwards-Anderson type order parameter) and the frequency dependent thermal conductivity. Using the RSRG-X, we find that in the vicinity of the critical temperature both observables show scaling behavior consistent with the infinite randomness critical point. The scaling behavior becomes non-analytic in the infinite system-size limit. Finally, we discuss the implications of our results in Sec.~\ref{sec:conclusions}.

\section{Dynamical phase diagram of the hJJ' model}
\label{sec:transition}

Our task is to develop a renormalization group procedure suited for describing the excited states of the weakly-interacting hJJ' model. Our approach is based on the core ideas of the RSRG methods developed for the ground state behavior of random magnets~\cite{Dasgupta1980,Bhatt1982,Fisher1992,Fisher1994,Motrunich2001}, and for the TFIM in particular~\cite{Fisher1995}. In the ground state methods, at each RSRG step we diagonalize the strongest local term of the Hamiltonian (\ref{hjjp}) and fix the corresponding subsystem in the ground-eigenstate of that term. We generate effective couplings between the subsystem and its neighboring spins through second-order perturbation theory, allowing only virtual departures from the chosen ground state of the strong term. In the TFIM, when the nearest-neighbor $x$-$x$ interaction is dominant, the RSRG steps lead to the formation of a macroscopic ferromagnetic cluster that breaks the Ising symmetry. On the other hand, when the transverse field is dominant, it pins most local spins to the $+z$ direction. The addition of weak interactions, $J'_i\ll J_i,h_i$ results in a shift of the critical point, but cannot affect the universal zero temperature properties of the TFIM.

\subsection{RSRG-X}

\begin{figure*}
\includegraphics[width=\textwidth]{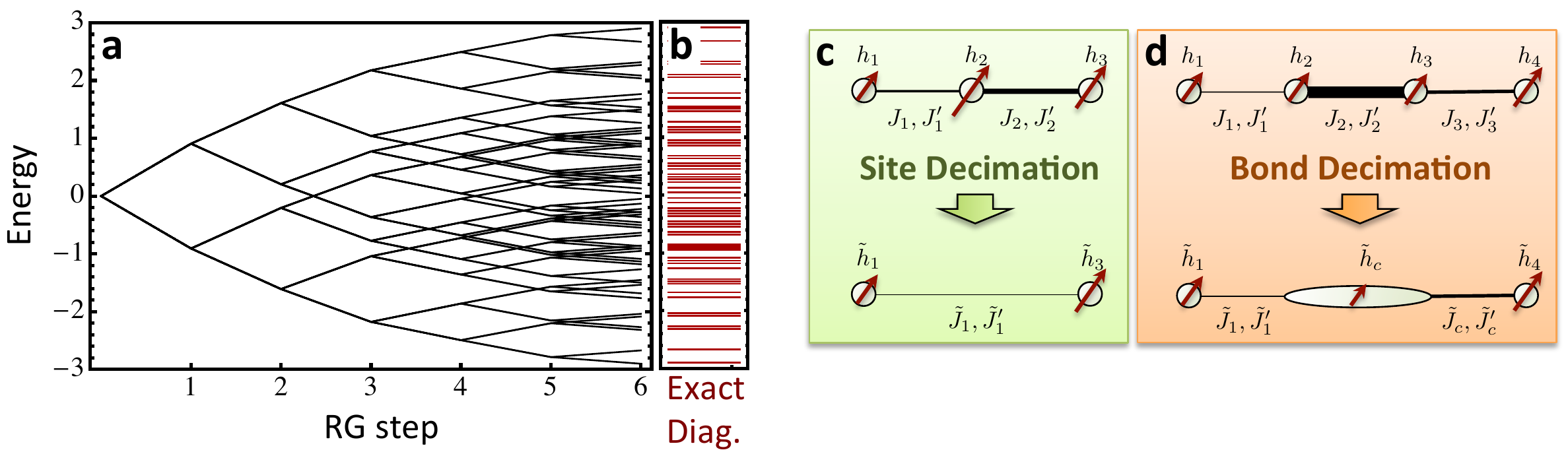}
\caption{
{\bf a} The RSRG-X tree for a six site Ising chain. The leaves of the tree correspond to the many-body eigenstates of the spin chain. {\bf b} The corresponding eigenspectrum found by exact diagonalization.
{\bf c} ({\bf d}) RSRG-X rules for site (bond) decimation in the hJJ' model. In the site decimation rule a local magnetic field ($h_2$) is the dominant coupling thus the corresponding local spin is either aligned (ground state) or anti-aligned (excited state) with it. Eliminating $h_2$, we obtain a new set of couplings $\tilde{h}_1$, $\tilde{J}_1$, $\tilde{J}_1'$, and $\tilde{h}_3$ to second order (see the methods section for details). In the bond decimation rule $J_2$ is the dominant coupling, corresponding to the neighboring spins being either aligned (ground state) or anti-aligned (excited state, i.e. a domain-wall) along the $x$-axis. 
\label{fig:two}
}
\end{figure*}

The crucial difference between our RSRG-X method and the ground-state RSRG methods has to do with the choice of the local eigenstates at each RG step. Instead of retaining only the lowest energy states at each RG step, we can choose either the low energy or the high energy manifold of the local term. The two manifolds are separated by a large gap, which controls the perturbation theory with which we generate the effective couplings. Thus, each RSRG-X step corresponds to a binary branching of a tree as illustrated in Fig. \ref{fig:two}a, where the leaves of the tree correspond to the actual many-body spectrum, Fig. \ref{fig:two}b.

Implementing the RSRG-X on the hJJ' model requires two types of RG decimation steps: site and bond decimations. Consider the $n$-th RSRG-X step. If the largest gap in the system is due to a field $h_2$ (see Fig.~\ref{fig:two}c), the site could be eliminated by letting spin $2$ be either mostly aligned ($c_n=+1$) or mostly anti-aligned ($c_n=-1$) with the local field. Second-order perturbation theory induces an effective Ising coupling between sites $1$ and $3$: $\tilde{J}\rightarrow c_n J_1 J_2/h_2$. Additionally, $J'$ couplings shift the local fields $\tilde{h}_{1(3)}\rightarrow h_{1(3)} + c_n J'_{1(2)}$. The significance of these shifts is that subsequent RSRG-X steps depend on the history of branching $c_n$'s at all higher energy scales. The history dependence is a feature of the interacting model and would not arise in the case $J'_i=0$.
If the largest gap is due to a bond $J_2$ (see Fig.~\ref{fig:two}d), the two sites $2$ and $3$ form a cluster which is either ferromagnetic ($c_n=+1$) or antiferromagnetic ($c_n=-1$). The rules for cluster formation also follow from second order perturbation theory, see methods for details. Due to our choice of the initial distributions, throughout the RSRG-X flow all $J'$ values are small compared to the nearby $J$ and $h$ values, and therefore, we do not implement a $J'$ elimination step. Once we have eliminated the largest term in the Hamiltonian, we reduce the scale of the largest gap, and then seek the next largest gap, until all degrees of freedom in the chain are eliminated.  

Obtaining the entire many-body spectrum requires the construction of all branches of the tree, which is an exponentially hard task. We therefore perform thermal sampling of typical eigenstates using Monte Carlo. In our procedure, Monte Carlo states correspond to the many-body eigenstates, which we describe via a sequence of branching choices $b=\{c_1, c_2, \dots c_L\}$ going from the root to a leaf of the tree, where $L$ is the number of sites in the chain. To obtain a new Monte Carlo state $b'$, we start at the leaf of the tree, go up the tree a random number of nodes, e.g. $m$, and flip $c_{L-m}\rightarrow -c_{L-m}$. Next, we calculate the energy change of the many-body state due to the flip of $c_m$, by performing the RSRG-X steps from $m$ to $L$, and accept or reject the new state $b'$ according to Metropolis~\cite{Metropolis1953}. 


A possible concern regarding the RSRG-X method is that it fails to correctly account for resonant coupling between nearly degenerate but distinct branches, which could result in delocalization. Ref.~\onlinecite{Vosk2013} has argued that in the presence of strong disorder these resonances are not strong enough to lift the near degeneracy of the eigenstates, hence our system always remains localized.

\subsection{RSRG-X flows}

The dynamical phase diagram of the hJJ' model derives from the flow of the full distributions of coupling constants $h$, $J$ and $J'$. 
We can, however, characterize the flow by tracking just two parameters: ${\cal C}= \llbracket \langle \text{avg}[\ln |h|] - \text{avg}[\ln |J|] \rangle_\text{Th} \rrbracket$ and ${\cal D}=\llbracket \langle \text{var}[\log |h|] +\text{var}[\log |J|] \rangle_\text{Th} \rrbracket$, where 
$\langle\ldots \rangle_\text{Th}$ and $\llbracket \ldots \rrbracket$ represents thermal and disorder averaging. Because $J'$ is irrelevant in the RG sense, we do not explicitly characterize the flow of the $J'$ distribution. ${\cal C}$ measures the relative strength of $h$ and $J$ distributions: ${\cal C}>0$ indicates that $h$'s are dominant and the flow is towards the paramagnetic phase while ${\cal C}<0$ indicates that $J$'s are dominant and the flow is towards the Hilbert-glass. ${\cal D}$ measures the disorder strength, which always grows upon coarse graining. In the $J'=0$ case, the ratio $\delta={\cal C}/{\cal D}$ remains almost constant throughout the RG flow~\cite{Fisher1995}, and hence we use the initial value of $\delta_I$ as a tuning parameter.

A peculiar property of the interaction-free TFIM (i.e. $J'_i=0$) is that the absolute values of the effective couplings following any decimation are insensitive to the choice of final state (i.e., independent of $c_n$'s). Therefore, the fate of the RSRG-X flow is solely determined by $\delta_I$. Hence, the model experiences a dynamical transition at a fixed critical value of $\delta_I$ independent of temperature: $\delta_I>0$ results in $h$ dominated flows towards the paramagnet while $\delta_I<0$ results in $J$ dominated flows towards the Hilbert-glass. 

\begin{figure*}
\includegraphics[width=\textwidth]{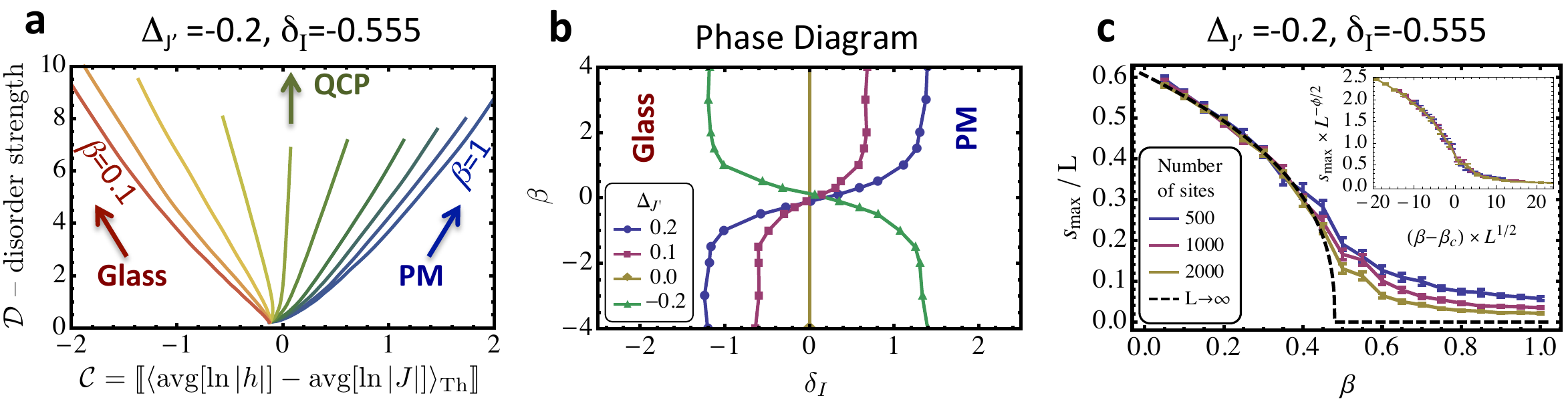}
\caption{{\bf a} Renormalization group flows in the ${\cal C}$ -- ${\cal D}$ plane, where ${\cal C}$ measures whether local fields or bonds are stronger while ${\cal D}$ measures the strength of disorder. 
Tuning $\beta$ results in RSRG-X flows towards the Hilbert-glass phase, the quantum critical point, and the paramagnetic phase as indicated by arrows.
[the initial value of the tuning parameter $\delta$ and the width of the initial $J'$ distribution $\Delta_{J'}$ is held fixed, see text]. 
{\bf b} Phase diagram of the hJJ' model in the $\delta$ -- $\beta$ plane for several values of $\Delta_{J'}$.
{\bf c} Size of the largest cluster $s_{\text{max}}$ as a function of $\beta$ for several values of the system size $L$. The transition becomes sharper and sharper as $L$ increases. The inset shows a finite-size scaling collapse of the same data. (Error bars indicate uncertainty due to disorder averaging, uncertainty due to Monte Carlo sampling is always smaller.) \label{clustersize}
\label{fig:three}}
\end{figure*}

When interactions are added, temperature starts playing an important role because gaps produced by decimation steps are no longer independent of the choice of eigenstate.  Thermal averaging biases decimation steps towards low or high energies, when $\beta=1/T$ is positive or negative respectively. This leads to a temperature-driven dynamical transition, which will be the focus of this work. The transition is encoded in the flows shown in Fig. \ref{fig:three}a: fixing $\delta_I$ we observe that the inverse temperature $\beta$ tunes whether the flow is towards the Hilbert-glass or the paramagnet. 

\subsection{Dynamical phase diagram}
The flows obtained from the RSRG-X method reveal a dynamical phase diagram of the hJJ' model. Fig. \ref{fig:three}b displays several cross sections of the phase diagram in the space of detuning versus inverse-temperature for various values of the interaction strength.
We describe the strength of the interactions by the parameter $\Delta_{J'}$, which measures the width of the $J_i^{'}$ distribution;  $\Delta_{J'}>0$  ($\Delta_{J'}<0$) corresponds to the initial distribution with all $J_i^{'}>0$ ($J_i^{'}<0$). As $\Delta_{J'}$ (and hence typical $J'$) decreases, the sigmoid phase transition curve narrows, becoming vertical at the non-interacting point $J'=0$ and finally mirroring itself for negative values of $\Delta_{J'}$.

The shape of the phase diagram can be understood by noting the effect of the site decimation RG rule on the initial distributions (we always use $J_i>0$, $h_i>0$). When $\Delta_{J'}>0$, $J'$ increases $h$ on neighboring sites in the excited states branches and reduces it in the low-energy branches. Tuning the temperature changes the branching ratios which effectively alters the tuning parameter $\delta_I$. Hence, we expect the shape of the transition line to be $\delta_I \approx C_0 \Delta_{J'} \tanh(E_\text{char}/T_c)$, where $T_c$ is the critical temperature, $E_\text{char}$ is the characteristic energy scale describing the initial $J$ and $h$ distributions and $C_0$ is a constant related to the shape of the initial $J'$ distribution. With the exception of a small offset, this shape is indeed observed in Fig. \ref{fig:three}b.

The offset can be understood from the infinite temperature limit: $\beta\rightarrow 0$ brings about an equal superposition of all branches which washes away any $J'$ effects. The remaining $J'^2$ effects, associated with the flow of $\langle \log h \rangle$, displace the transition line from the origin:
\be
\delta_{I,c}(\beta\rightarrow 0) = C_0 E_\text{char}\beta_c\Delta_{J'} + C_2 \Delta_{J'}^2,
\ee
where $C_2$ is also constant. 

\section{Dynamical spin correlations and the order parameter}
\label{sec:autocorrelation}

Armed with the RSRG-X method we can now characterize the two dynamical phases and the phase transition between them. In the Hilbert-glass phase, the many-body eigenstates tend to contain an infinite cluster that is produced by bond-decimations. We can crudely think of these eigenstates as having a random sequence of fixed values of $\sigma^x$ on the different sites belonging to the infinite cluster. Thus, each eigenstate breaks the $\mathbb{Z}_2$ symmetry of the Ising model. Of course, different eigenstates, even very close in energy, typically have completely different spin configurations. 

A natural signature of the cluster phase is the dynamic spin autocorrelation function 
\be
\llbracket \gamma (t) \rrbracket \equiv \left\llbracket \summ_n\frac{e^{-E_n/T}}{Z}\bra{\psi_n} \sx_i(t)\sx_i(0)\ket{\psi_n} \right\rrbracket.
\ee
For an infinite system in the cluster phase we expect this correlator to saturate: 
$\lim_{t\rightarrow \infty} \llbracket \gamma(t) \rrbracket \rightarrow f$,
where $f=s_{\text{max}}/L$ is the ratio of the number of sites participating in the
largest cluster over the bare size of the chain~\footnote{In a finite system, the spins belonging to the largest cluster can collectively flip on an exponentially long timescale set by the system size. Therefore, one has to consider the analogous four-point correlator 
\unexpanded{$
\llbracket \langle 
\sx_i(t)\sx_i(0)\sx_j(t)\sx_j(0)
\rangle_\text{Th} \rrbracket
$}.}. Alternatively, the glass order parameter introduced by Edwards and Anderson (EA) can capture this kind of static eigenstate order and moreover survive thermal averaging
 \be
m_\text{EA} \equiv \left\llbracket \summ_n\frac{e^{-E_n/T}}{Z}|\bra{\psi_n} \sx_i\ket{\psi_n}|^2 \right\rrbracket.
\ee
It is important to note that although the EA order parameter offers a static signature of the phase it is not a thermodynamic observable~\footnote{Due to the squared expectation value, $m_\text{EA}$ cannot be written in the form $\mathrm{tr}\left(\rho A\right)$, where $\rho$ is the equilibrium density matrix and $A$ is an operator} and hence does not entail a thermodynamic singularity.


To quantify the transition, we have used RSRG-X to directly measure the fraction of the system occupied by the largest cluster $s_{\text{max}}/L$. Fig. \ref{fig:three}c shows the evolution of the largest cluster fraction across the phase transition, as we tune the temperature. As the system size increases, the transition becomes sharper, approaching the $L\rightarrow \infty$ asymptote indicated by the dashed line. Indeed, this notion can be made concrete using a finite-size scaling hypothesis
\begin{align}
s_{\text{max}}=L^{\phi/2} f \left[ (\beta-\beta_c)^{1/\nu} L \right],
\end{align}
to successfully collapse the $s_{\text{max}}$ data (see Inset of Fig. \ref{fig:three}c). To obtain this hypothesis, we used the intuition that temperature ($\beta$) tuning is equivalent to $\delta_I$ tuning combined with two zero temperature results: (1) correlation length critical exponent $\nu=2$ and (2) at criticality $s_{\text{max}}\sim L^{\phi/2}$ (where $\phi$ is the Golden Ratio)~\cite{Fisher1994}. 

\section{Localization and heat conductivity}
\label{sec:sigmaT}
Transport measurements are an important experimental tool that could be used for the detection of dynamical phase transitions. Here, we present the first calculation of heat transport using RSRG methods. We focus on the heat transport since the energy current is the only locally conserved current in the $hJJ'$ model. It is important to note that both the Hilbert-glass and the paramagnetic phases are many-body localized. They do not thermalize by themselves and feature strictly vanishing DC heat conductivity. Nevertheless, it is interesting to characterize these phases by their AC transport in the low frequency limit.

\begin{figure}
\includegraphics[width=7cm]{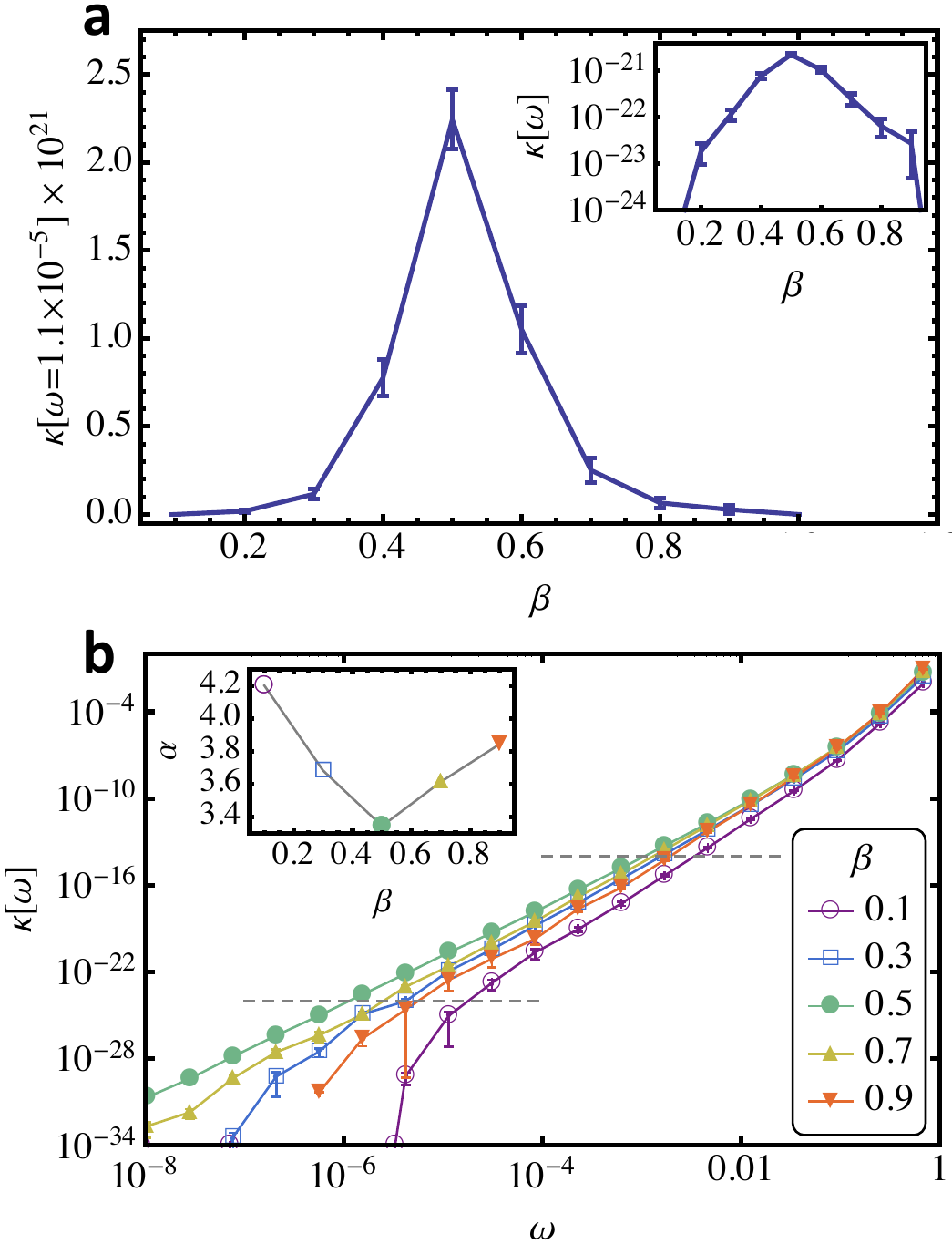}
\caption{{\bf a} Low frequency heat conductivity plotted as a function of temperature. At the phase transition the system becomes least localized, which is associated with a cusp in the heat conductivity. Inset: same data plotted on a log-linear scale supporting the form $\kappa \sim e^{-\text{const} |\beta-\beta_c|}$. {\bf b} Heat conductivity plotted as a function of frequency for five different temperatures. The low frequency heat conductivity displays power law scaling with frequency $\kappa(\omega) \sim \omega^\alpha$. Inset: $\alpha$ as a function of $\beta$ (see text). (In all cases $\delta_I=-0.555$ and $\Delta_{J'}=-0.2$. Error bars indicate uncertainty due to disorder averaging, uncertainty due to Monte Carlo sampling is always smaller. Dashed lines in {\bf b.} indicate the region being fitted for the inset.)
\label{fig:four}}
\end{figure}

The calculation of the heat conductivity for the hJJ' chain is a potent demonstration of the RSRG-X's strength. Using the detailed knowledge of the local decimation steps leading to each branch in the many-body state tree, we can construct the Kubo integral (see Methods) which yields the real part of the conductivity:
\be
\kappa(\omega)=\frac{1}{i\omega L}\summ_{i,j}\int_{-\infty}^{0}dt e^{-i\omega
  t}\langle \l[ j_i(t),j_j(0)\rr]\rangle_\text{Th}.
  \label{kappa}
\ee
Here, $j_i(t)=\l[\summ_{k=1}^{i-1}\H_i,\H\rr]$ is the energy current operator and $\H_k$ the Hamiltonian pertaining to site $k$ and the bond $k$, $k+1$.  Plotting the low frequency heat conductivity as a function of $\beta$ (see Fig. \ref{fig:four}a), we observe a cusp-like feature signaling the onset of the Hilbert-glass transition. 

To understand the origin of this feature, and how it becomes sharper as $\omega \rightarrow 0$, we  consider the scaling of $\kappa(\omega)$. In the non-interacting case, subsequent RSRG-X steps are independent of the preceding steps, and hence we can obtain (see methods) 
\begin{align}
\kappa(\omega) \sim \omega^\alpha.
\end{align} 
At criticality, $\alpha=3$, which is the engineering dimension of $\kappa(\omega)$. As we tune away from criticality, we observe the appearance of an anomalous dimension $\alpha \sim 3 + \text{const} \times |\delta_I|$ (up to logarithmic corrections, see methods). As $J'$ is an irrelevant operator, we expect this scaling to persist in the presence of $J'$. Furthermore, using the equivalence of $\beta$ and $\delta_I$ tuning, we hypothesize that $\alpha \sim 3 + \text{const} \times |\beta-\beta_c|$ when we tune across the transition using temperature. 

We test our scaling hypothesis in two ways. First, we plot $\kappa$ for a small fixed $\omega$ as a function of $\beta$ on semi-log axis (see inset of Fig.~\ref{fig:four}a). We observe essentially linear behavior away from the transition that is consistent with the hypothesis. Second, in Fig.~\ref{fig:four}b, we plot $\kappa[\omega]$ as a function of $\omega$ for several values of temperature. $\kappa[\omega]$ displays non-universal behavior at high frequencies, which becomes a power law at low frequencies, and is cut-off by the system size at very low frequencies. As we tune $\beta$ towards the transition from either side, we observe that the power-law exponent $\alpha$ decreases, reaching its lowest value of $3.3$ at the critical point (see inset of Fig.~\ref{fig:four}b). Thus, up to logarithmic corrections that likely account for the deviation away from the engineering value $\alpha=3$ in our $2000$ site spin chain, we find that our numerical results are consistent with the expected critical scaling at the transition.

\section{Conclusions}
\label{sec:conclusions}
We have explored the notion of a dynamical finite temperature transition by investigating a concrete model: the hJJ' model. This investigation is made possible by the development of the real space renormalization group method for excited states (RSRG-X), which we also describe in this manuscript. Although notions of a dynamical finite temperature phase transition have appeared in the literature before, we offer a detailed description of a specific example of this type of transition. 

Specifically, in the thermodynamic sense the hJJ' model has a $0^+$ and a $0^-$ temperature quantum phase transition. Away from these two zero temperature phase transitions, the model displays no singularities in any of its thermodynamic observables. However, connecting this pair of QCPs, we find the Hilbert-glass transition (HGT) line, associated with singular behavior in dynamical observables including dynamical spin auto-correlations and heat conductivity. We therefore add a new type of universality to the lore of critical phenomena.

Finally, we note that the RSRG-X method could find applications in describing many-body eigenstates, especially localized eigenstates, in other models both in 1D and in higher dimensions.

\section{Notes}
\begin{enumerate}
\item
During the preparation of the manuscript, a paper describing a method that has some similarities to our RSRG-X appeared on the arXiv~\cite{Swingle2013}.
\item
Another manuscript posted in parallel to this one explores the HGT from the complementary viewpoint of the time evolution following a quench~\cite{Vosk2013a}.
\end{enumerate}

\section{Acknowledgements}
It is our pleasure to thank Kedar Damle, Oleksei Motrunich, David Huse, and Stefan Kehrein for useful conversations. 
The authors thank the KITP and the National Science Foundation under Grant No. NSF PHY11-25915 for hospitality during the conception of the paper and the Aspen Center for Physics and the NSF Grant \#1066293 for hospitality during the writing of this paper. 
The computations in this paper were run on the Odyssey cluster supported by the FAS Science Division Research Computing Group at Harvard University.
The authors acknowledge support from the Lee A. DuBridge prize postdoctoral fellowship (DP), the IQIM an NSF center supported in part by the Moore foundation (DP \& GR), DMR-0955714 (VO), the Packard foundation (GR), BSF (EA \& ED), ISF and the Miller Institute for Basic Science (EA), Harvard- MIT CUA, the DARPA OLE program, AFOSR MURI on Ultracold Molecules,
and ARO-MURI on Atomtronics (ED).

\begin{widetext}
\section{Methods}
\label{sec:methods}

\subsection{RSRG-X general framework}
Our results rely on the extension of the real space renormalization group to obtaining and sampling the excited spectrum of a 1D problem. The method, which we refer to as RSRG-X, follows closely the ground-state relevant constructions in Refs.~\cite{Dasgupta1980, Bhatt1982, Fisher1992, Fisher1994, Fisher1995}. In both cases we iteratively eliminate the strongest local pieces of the hamiltonian, and construct the many-body state as an approximate tensor product of eigenstates of the strong terms. New low-energy terms arise during the elimination process, and they are calculated within second-order perturbation theory via a Schrieffer-Wolff transformation.  The method concludes when all degrees of freedom are eliminated, or when the Hamiltonian consists only of mutually commuting terms.

Let us describe a single iteration step of the RSRG-X method. Our analysis assume local Hamiltonians of the following form:
\be
\H=\summ_i \H_{i/2}
\label{conv}
\ee
where the terms with an integer index pertain to a site, and the terms with half-integer index pertain to bonds, and have a nontrivial effect on the tensor product of the two Hilbert spaces of the appropriate two neighboring sites. The first task in the RSRG-X elimination step is to separately solve the local Hamiltonians $\H_{i/2}$, and find the term $\H_{m/2}$ with the largest energy gap in its spectrum. 

Next, we concentrate on the three pieces $\H_{m/2,(m \pm 1)/2}=\H_{\frac{m-1}{2}}+\H_{m/2}+\H_{\frac{m+1}{2}}$, and partially solve for the many-body ground state using a Schrieffer-Wolff transformation. We assume that the largest gap separates the energies of two subspaces of the domain of $\H_m$; we denote them with the letters $a$ and $b$, and also define $P_a$ and $P_b$ as projections onto them. For convenience we rewrite the three-term Hamiltonian in the following way:
\begin{align}
\H_{m/2,(m\pm 1)/2}=H_0 + V.
\end{align}
$\H_0$ encodes the largest gap we identified, and ignores all the degeneracy breaking effects within the high and low energy subspaces. Without loss of generality, we write $H_0$ as:
\be
H_0=\frac{\lambda}{2} \left(
\begin{array}{cc}
I_a & 0\\
0 & -I_b
\end{array}
\right)
\ee
where $\lambda$ is the size of the largest local gap, and $I_a$ and $I_b$ are two identity matrices with dimensions corresponding to the dimension of the subspaces $a$ and $b$. Our goal is to find a unitary transformation $e^{iS}$, such that 
\begin{align}
\l[e^{i S} (H_0+V) e^{-i S},\,H_0\rr]=0, \label{eq:t1}
\end{align}
i.e., that eliminates the pieces in $V$ that mix the subspaces $a$ and $b$. We perform this task perturbatively, expanding  $S=S^{(0)}+S^{(1)}+S^{(2)}+\ldots$ in powers of $V$. 
 
To facilitate the task we introduce the notation that $P_\alpha$ is a projection onto subspace 
$\alpha \in \{ a,b \}$, $P_{\bar{\alpha}}$ is the projection onto the complimentary subspace, $(-1)^a=1$ and $(-1)^b=-1$. $S^{(0)}=0$ since $H_0$ does not connect the two subspaces. $S^{(1)}$ is found by solving $P_\alpha \left([i S^{(1)},H_0]+V\right) P_{\bar{\alpha}}=0$, which yields
\begin{align}
S^{(1)}=\frac{1}{i\lambda} \sum_\alpha \frac{(-1)^\alpha}{i\lambda} P_\alpha V P_{\overline{\alpha}}.
\end{align}
We note that due to the structure of $H_0$, $[i S^{(1)},H_0]+\sum_{\alpha} P_\alpha V P_{\overline{\alpha}}=0$. Using $S^{(1)}$ we can find the effective Hamiltonian, for the two subspaces, up to second order
\be
\H_\text{eff}=e^{i S}(H_0+V) e^{-i S} - H_0 \approx \summ_{\alpha=a,\,b}P_{\alpha} \l(V+ \l[i S^{(1)},V\rr]-\frac{1}{2}\l\{(S^{(1)})^2,H_0\rr\}+S^{(1)}H_0 S^{(1)}\rr)P_{\alpha},
\ee
where $\{f,g\}$ represents the anti-commutator $fg+gf$.

If we are interested in the corrections to the wave functions, we need to know the unitary transformation $e^{iS}$ up to second order.  $S^{(2)}$ is found by solving $P_\alpha \left( [i S^{(2)},H_0]+[i S^{(1)}, V] - \frac{1}{2} \{(S^{(1)})^2, H_0  \}+ S^{(1)} H_0 S^{(1)}  \right) P_{\bar{\alpha}}=0$, which yields
\begin{align}
S^{(2)}=\frac{1}{i\lambda^2} \sum_\alpha P_\alpha V (P_\alpha-P_{\overline{\alpha}}) V P_{\overline{\alpha}}.
\end{align}

To make the RSRG-X method work, the Schrieffer-Wolff perturbation theory must converge. We infer from the above forms that a necessary condition is that the largest eigenvalue must be smaller in absolute value than $\lambda$: $||V||\ll\lambda$. This should also be sufficient in most physical cases, but we do not attempt to pursue a proof of this notion, but rather rely on small-size numerics to confirm the validity of the method. 

\subsection{Application to the hJJ' model}
The hJJ' naturally fits the mold of the RSRG-X, just as the TFIM has become the ubiquitous example for the ground state RSRG method. Following the convention of Eq. (\ref{hjjp}), we write the site and bond terms, respectively, as:
\begin{align}
\H_{i}&=-h_i\sz_i,&\H_{i-1/2}&=-J_i\sx_i\sx_{i+1}-J'_i\sz_i\sz_{i+1}.
\end{align}
We concentrate on the case in which $J'$ is small and therefore can be treated perturbatively. In this case, the largest gap in the chain may arise either due to a site Hamiltonian, i.e., a large field $h_i$, or a bond Hamiltonian, i.e. a large Ising term $J_i$. For notational simplicity, we assume that the largest coupling occurs on site $i=2$, and we refer to Fig.~\ref{fig:two}c \& d for the definition of the notation.  Applying the Schrieffer-Wolfe transformation, we find the site dominant rule to be
\begin{align}
\tilde{h}_1&\rightarrow h_1 + c J_1', & \tilde{h}_3&\rightarrow h_3 + c J_2',\\
\tilde{J}_1&\rightarrow c \frac{J_1 J_2}{h_2}, & \tilde{J}_1&'\rightarrow 0 \\
\lambda &\rightarrow c \left(h_2+\frac{J_1^2+J_2^2}{2h_2}\right),
\end{align}
and the bond dominant rule to be
\begin{align}
\tilde{h}_1&\rightarrow h_1 + c \frac{h_2 J_1'}{J_2}, & \tilde{h}_c&\rightarrow J_2'+c\frac{h_2 h_3}{J_2}, & \tilde{h}_4&\rightarrow h_4 + c \frac{h_3 J_3'}{J_2},\\
\tilde{J}_1&\rightarrow c J_1,& \tilde{J}_c&\rightarrow J_3,\\
\tilde{J}_1'&\rightarrow c\frac{h_3 J_1'}{J_2},& \tilde{J}_c'&\rightarrow c\frac{h_2 J_3'}{J_2}\\
\lambda &\rightarrow c \left(J_2+\frac{h_2^2 + h_3^2+(J_1')^2+(J_3')^2}{2J_2}\right).
 \end{align}
Here, $c=\pm1$ indicates the choice of the local Hilbert space (tree branch) and $\lambda$ is the shift of the energy eigenvalue.

The most important thing in the RSRG-X procedure is that no relevant new terms are produced, and the decimation step could repeat itself until all degrees of freedom in the chain are accounted for. 

\subsection{Numerics}
The RSRG-X procedure was performed numerically on random spin chains containing from $500$ to $2000$ spins. For each realization of disorder, we performed $5 \times L$ equilibration Monte Carlo steps, followed by $2 \times 10^5$ to $10^6$ data collection Monte Carlo steps. The initial distributions for $J_i$, $h_i$, and $J_i'$ were selected to be uniform distributions with the following ranges: 
\begin{align}
P[J]&=U(0.3,1-s), & P[h]&=U(0.3,1+s), & P[J']&=U(0,\Delta_{J'})
\end{align}
were $U[a,b]$ indicates the uniform distribution that ranges from $a$ to $b$, $s$ is the initial shift of the distribution (which we translate into the initial value of the tuning parameter $\delta_I$), and $\Delta_{J'}$ corresponds to the width of the initial $J'$ distribution. We note that while we only used positive definite initial distributions of $J$ and $h$, we looked at both positive-definite ($\Delta_{J'}>0$) and negative-definite ($\Delta_{J'}<0$) initial distributions of $J'$.

\subsection{Derivation of the heat conductivity}
To find the heat current operator we apply the relation $j_i=\l[\summ_{k=1}^{i}\H_i,\H\rr]$ to the TFIM model and obtain $j_i=-2 i J_i h_{i+1} \sigma^x_i \sigma^y_{i+1}$. Here, we do not include $J'$ contributions because $J'$ is an irrelevant operator which will not affect the low frequency thermal conductivity. In order to compute the heat current operator at a given RG step, we use the renormalized Hamiltonian in the commutation relation. We note that using the Schrieffer-Wolff transformation on the heat current operator as we run the RG yields exactly the same result as the direct calculation using the renormalized Hamiltonian because the heat current operator is directly tied to the Hamiltonian.

To compute the thermal conductivity, we follow Ref.~\onlinecite{Motrunich2001} and write the Kubo formula Eq.~\ref{kappa} in terms of an explicit trace over many-body eigenstates
\begin{align}
\kappa(\omega) = \frac{\pi}{\omega Z L} \sum_{m,n} \delta(\omega+E_m-E_n) e^{-\beta E_m} (1-e^{-\beta \omega}) |\langle n | \sum_i j_i | m \rangle|^2,
\end{align}
where $Z$ is the partition function. To evaluate $\kappa$ using Monte Carlo, we sample over the states $|m\rangle$. To determine $E_n$ and the matrix element $|\langle m | \sum_i j_i | n \rangle|$, we assume that the most important contribution comes from the local Hamiltonian that is being decimated at the scale $\omega$. That is, if we are decimating the bond $J_k$, then $E_n=E_m-\omega \approx E_m - 2 J_k$ and $|\langle m | \sum_i j_i | n \rangle| \approx 2 J_k h_{k+1}$. Similarly, if we are decimating the site $h_k$, $E_n = E_m - 2 h_k$ and $|\langle m | \sum_i j_i | n \rangle| \approx 2 J_{k-1} h_k$.

In the non-interacting case, we can evaluate $\kappa$ explicitly, as the partition function becomes a product of partition functions of the individual RG steps. Thus, we can explicitly write down the heat conductivity
\begin{align}
\kappa(\omega) = \frac{1}{\omega} n(\omega) P(\omega; \omega) \int_0^\omega R(\omega'; \omega) (2 \omega \omega')^2 d\omega \, \tanh\left(\frac{\omega}{T}\right) + (P \leftrightarrow R),
\end{align}
where $n(\omega)$ is the fraction of the sites that survive at the scale $\omega$ and $P(\cdot,\omega)$ [$R(\cdot,\omega)$] is the distribution of $|J|$'s [$|h|$'s] at scale $\omega$. Using the critical and off-critical distributions from Ref.~\onlinecite{Fisher1995}, we find
\begin{align}
\kappa(\omega)&\sim 
\left\{
\begin{array}{cc}
\omega^{3+2|\delta|}&\omega<T,\\
\omega^{2+2|\delta|}&\omega>T.
\end{array}
\right.
\end{align}
Exponents at criticality match expectations from the engineering dimensions: using the dimensionality of the heat current operator $[j_k(\omega)]\sim\omega^2$, the distribution functions $[P(\cdot;\omega)]=[R(\cdot;\omega)]=\omega^{-1}$, and remembering that $\tanh\left(\frac{\omega}{T}\right)$ contributes a power of $\omega$ in the `hydrodynamic' regime $\omega \ll T$. The first logarithmic corrections to this result is $\log(1/\omega)^{-4}$, where two powers of the logarithm come from the density of states, one power comes from the $J$ distribution and one power from the $h$ distribution.

\end{widetext}

\bibliography{dp}

\end{document}